\documentclass[dvips,12pt]{article}
\usepackage{graphicx}
\renewcommand{\baselinestretch}{1.5} 
\begin{document} 
\thispagestyle{empty} 
\begin{center}
{\large{\bf
A numerical algorithm for efficiently obtaining a
Feynman parameter representation
of one-gluon loop QCD Feynman diagrams
for a large number of external gluons
\\}}
\vspace{1cm} 
{\large A.~S.~Kapoyannis*, A.~I.~Karanikas and C.~N.~Ktorides}\\ 
\smallskip 
{\it University of Athens, Physics Department\\
Nuclear \& Particle Physics Section\\
Panepistimiopolis, Ilisia GR 157--71, Athens, Greece}\\
\vspace{1cm}

\end{center}
\vspace{0.5cm}
\begin{abstract}

A numerical program is presented which facilitates a computation pertaining to
the full set of one-gluon loop diagrams (including ghost loop
contributions), with M attached external gluon lines in all possible
ways. The feasibility of such a task rests on a suitably defined master
formula, which is expressed in terms of a set of Grassmann and a set of
Feynman parameters.
The program carries out the Grassmann integration and performs the Lorentz
trace on the involved functions, expressing the result as a compact sum of
parametric integrals. The computation is based on tracing the structure of
the final result, thus avoiding all intermediate unnecessary calculations
and directly writing the output.
Similar terms entering the final result are grouped together. The running time
of the program demonstrates its effectiveness, especially for large M.
\end{abstract}

\vspace{2cm}


*Corresponding author.
{\it E-mail address:} akapog@phys.uoa.gr (A. S. Kapoyannis)

\newpage
\setcounter{page}{1}

{\large \bf PROGRAM SUMMARY}

{\it Manuscript title:} A numerical algorithm for efficient computations
of one-gluon loop Feynman diagrams in QCD for a large number of external
gluons

{\it Authors:} A.~S.~Kapoyannis, A.~I.~Karanikas and C.~N.~Ktorides

{\it Program title:} DILOG2

{\it Programming language:} FORTRAN 90

{\it Computer(s) for which the program has been designed:} Personal Computer

{\it Operating system(s) for which the program has been designed:}
Windows 98, XP, LINUX


{\it Number of processors used:} one  
 


{\it CPC Library Classification:} 11.5 Quantum Chromodynamics, Lattice
Gauge Theory

{\it External routines/libraries used:} none

{\it CPC Program Library subprograms used:} none


{\it Nature of problem:}
The computation of one gluon/ghost loop diagrams in QCD with many external gluon lines is a
time consuming task, practically beyond reasonable reach of analytic
procedures. We apply recently proposed master formulas towards
the computation of such diagrams with an arbitrary number
($M$) of external gluon lines, achieving a final result which reduces the
problem to one involving
integrals over the standard set, for given $M$, of Feynman parameters.

{\it Solution method:}
The structure of the master expressions is analysed from a numerical computation point of view. 
Using the properties of Grassmann variables
we identify all the different forms of terms that appear in the final result. Each form
is called ``structure''. We calculate theoretically the number of terms
belonging to every ``structure''. We carry out the calculation organising the
whole procedure into separate calculations of the terms belonging to every
``structure''. Terms which do not contribute to the final result are thereby
avoided. The final result, extending to large values of $M$, is also presented with terms
belonging to the same ``structure'' grouped together.
     
{\it Restrictions:} $M$ is coded as a 2-digit
integer. Overflow in the dimension of used array is expected to appear for
$M\geq20$ in a processor that uses 4-bytes integers or for $M\geq34$
in a processor with 8-bytes integers.

{\it Running time:} Depends on $M$, see enclosed figures.

\newpage

{\large \bf LONG WRITE-UP}

{\large \bf 1. Introduction}

Quantum Chromodynamics (QCD) enjoys universal acceptance as the 
fundamental theory for the strong interaction. As a quantum field theoretical
system, QCD has been extensively applied to situations in which its
perturbative content provides a dependable computational tool. It is, in fact,
within the framework of this perturbative content that QCD has successfully
confronted the quantitative description of the multitude of scattering
processes, which probe strong interaction dynamics at high energies.
Admittedly, the study of the non-perturbative domain of the theory offers
intriguing and, most certainly, fundamental challenges. Nevertheless, the
immediate need to confront recent measurements coming from the HERA and
Tevatron particle accelerators as well as the expected ones, in the near
future, from the LHC accelerator continues to put perturbative QCD (pQCD)
at the forefront of theoretical activity.

Given the non-abelian structure of QCD, the (by far) most demanding component of 
the theory, with respect to perturbative calculations, is its gluonic, as
opposed to its quark, sector\footnote{For that matter, this is more so the
case for the non-perturbative domain of the theory.}. In particular,
perturbative computations involving Feynman diagrams with gluon/ghost loops
become, to say the least, quite monstrous. During the last decade or so
various methods, aiming to expedite Feynman diagram computations in QCD, have
been proposed whose basic feature is that they rely in a first, rather than
the usual second, quantization approach to the formulation of the theory.
Corresponding attempts have employed either strings [1-3], or world-line
paths [4-11] as their underlying basic agents. Within the framework of the
latter case, two of the present authors [10,11], have been involved in work
which led to the formulation of a set of master expressions, that condense
the multitude of all Feynman diagrams entering a given configuration.
These expressions are determined by the number of loops and the number of
external gluon propagators attached to them. To be more precise, the derived
expressions  go up to two loop configurations, nevertheless the ``logic'' of
the construction can be extended to loops of higher order. Suffice it to say,
at this point, that the analytical confrontation of a two loop situation with
four ``external'' gluon lines constitutes a challenging enough problem [12].

The basic feature of the master expressions arrived at in [10,11] is that they
are furnished in terms of a set of Grassman and a set of Feynman variables.
Once integrations over these two sets of variables are performed one obtains
the full result, i.e. the one which, for the given configuration, contains
the contribution of all Feynman diagrams at once. It is obvious, even before
laying an eye on these master formulas, that in order to put them into
practical use, their confrontation calls for the employment of suitable
computational methods. It is the aim of this paper to present a program, which will be 
applied to the one gluon loop case for $M$ external gluonic lines
($M$ fairly large). The main part of our program deals with the
confrontation of multi-Grassmann variable integrals and arrives at
expressions which involve the appropriate set of Feynman parameters only.
This program could hopefully find applicability to other situations,
where multi-Grassmann variable integrals also make their entrance. 

Our paper is organized as follows. In the following section we present the
battery of formulas, which are associated with the master expression
corresponding to one gluon/ghost loop with $M$ external gluon attachments in
all possible ways. We shall consider cases up to fairly large values of $M$, which
exhibit divergent terms\footnote{As expected, the aforementioned master
expressions implicate the absence of divergent terms for $M>4$, cf.~[10].}, in
addition to finite ones.
In section 3 we present the ideas on which this program is based along with
the formulas which count the terms appearing in our
final result. In section 4 we describe the structure of the program, while
section 5 presents our results, accompanied by tables and figures.
Finally, our concluding remarks are made in section 6.

\vspace{0.5cm}
{\large \bf 2. The one loop master formula}

Consider a configuration consisting of one gluon/ghost loop onto which $M$ external 
gluon lines, with corresponding momenta $p_1\cdot\cdot\cdot p_M$ are attached (see 
Figure 1). According to [10], the master expression, which summarizes the total 
contribution from all Feynman diagrams pertaining to this configuration is given by
\[
\Gamma_1^{(M)}(p_1,\dots,p_M)=-\frac{1}{2} g^M (2\pi)^4
\delta^{(4)} \left(\sum_{n=1}^{M}p_n\right)
Tr_C (t_G^{\alpha_M} \cdots t_G^{\alpha_1}) \frac{1}{(4\pi)^2}
\int_0^\infty dT T^{M-3} \times
\hspace{10cm}\]
\[
\times \left[ \prod_{n=M}^1 \int_0^1 du_n \right]
\theta (u_M,\ldots,u_1) F^{(M)}(u_1,\ldots,u_m;T)
\exp\left[T\sum_{n<m} p_n \cdot p_m G(u_n,u_m) \right]
\hspace{10cm}\]
\begin{equation}
+ permutations \hspace{10cm}\;,
\end{equation}
where $g$ is the coupling constant of the theory, the
$t_G^{\alpha_i}$, $i=1,\dots,M$ are the
$SU(3)_{color}$ group generators (in the adjoint representation) with $Tr_C$
the trace over the color group, the $u_i$ are Feynman parameters, the
function $\theta$ is specified by
\[
\theta(u_M,\ldots,u_1)=\theta(u_M-u_{M-1})\ldots\theta(u_2-u_1)
\]
and
\[
F^{(M)}(u_1,\ldots,u_M;T)=
\left[\prod_{n=M}^1 \int d\xi_n d\bar{\xi}_n\right]
\left(Tr_L \Phi^{[1]} -2 \right) \times \hspace{5cm}
\]
\begin{equation}
\times \exp\left[\sum_n \sum_{m \neq n} \bar{\xi}_n \xi_n 
\varepsilon^n \cdot p_m \partial_n G(u_n,u_m)+
\frac{1}{2T} \sum_n \sum_{m \neq n} \bar{\xi}_n \xi_n \bar{\xi}_m \xi_m 
\varepsilon^n \cdot \varepsilon^m \partial_n \partial_m G(u_n,u_m)\right]\;.
\end{equation}

In the above relation the $\xi$'s are Grassmann variables,  the $\varepsilon^i$
are polarization vectors for the external gluons, $\Phi^{[1]}$ is the
so-called {\it spin factor} entering the world-line description of QCD (see
below), with $Tr_L$ denoting trace with respect to Lorenz generator
representation indices and the $G(u_n,u_m)$ are free propagators for the
particle modes entering the worldline path integral description of QCD, in
the context of its first quantized version (see [10]). They obey the
equation(s)
\begin{equation}
-\partial_n\partial_m G(u_n,u_m) = \partial_n^2\dot{G}(u_n,u_m) \equiv
\ddot{G}(u_n,u_m)=2[\delta(u_n,u_m)-1]\;,
\end{equation}
with boundary condition
\begin{equation}
\partial_n G(u_n,u_m) \equiv \dot{G}(u_n,u_m)=
sign(u_n-u_m)-2(u_n-u_m)=-\dot{G}(u_m,u_n)\;.
\end{equation}
The explicit expression for the spin factor in terms of the set of parameters
entering our expressions is (the $J_{\mu\nu}$ are the Lorentz generators, in
the vector representation)
\[
\Phi_{\mu\nu}^{[1]}=
P\exp\left[\frac{i}{2}\sum_{n=1}^{M} J \cdot \phi(n) \right]_{\mu\nu} =
\hspace{10cm}
\]
\begin{equation}
=\delta_{\mu\nu}+\frac{i}{2}(J_{\rho\sigma})_{\mu\omega}
\sum_{n=1}^{M}\phi_{\rho\sigma}(n)+
(\frac{i}{2})^2 (J_{\rho_2\sigma_2})_{\mu\lambda}
(J_{\rho_1\sigma_1})_{\lambda\nu}
\sum_{n_2=1}^M \sum_{n_1=1}^{n_2-1}
\phi_{\rho_2\sigma_2}(n_2) \phi_{\rho_1\sigma_1}(n_1) + \ldots \;,
\end{equation}
where
\begin{equation}
\phi_{\mu\nu}(n)=2\bar{\xi}_n\xi_n
(\varepsilon_\mu^n p_{n,\nu}-\varepsilon_\nu^n p_{n,\mu})
+\frac{4}{T} \bar{\xi}_{n+1}\xi_{n+1} \bar{\xi}_n\xi_n
(\varepsilon_\mu^{n+1} \varepsilon_\nu^n -
 \varepsilon_\nu^{n+1} \varepsilon_\mu^n ) \delta(u_{n+1}-u_n)\;.
\end{equation}

A point of note is the following: In the above expressions a specific time
ordering has been chosen according to which index $n+1$ comes immediately
after index $n$, with $\xi_{M+1}=\bar{\xi}_{M+1}=0$.

The saturation of indices $\rho$, $\sigma$ in $J_{\rho\sigma}$ is performed
instantly, since
\begin{equation}
(i/2)(J_{\rho\sigma})_{\mu\nu} 2
(\varepsilon_\rho^n p_{n,\sigma}-\varepsilon_\sigma^n p_{n,\rho})=
-2(\varepsilon_\mu^n p_{n,\nu}-\varepsilon_\nu^n p_{n,\mu})\;,
\end{equation}
\begin{equation}
(i/2)(J_{\rho\sigma})_{\mu\nu} (4/T)
(\varepsilon_\rho^{n+1} \varepsilon_\sigma^n-
 \varepsilon_\sigma^{n+1} \varepsilon_\rho^n)=
-(4/T)(\varepsilon_\mu^{n+1} \varepsilon_\nu^n-
 \varepsilon_\nu^{n+1} \varepsilon_\mu^n)
\end{equation}

Thus, eq.~(5) may be rewritten as
\begin{equation}
\Phi_{\mu\nu}^{[1]}=
\delta_{\mu\nu}+(-1) \sum_{n=1}^{M}\phi_{\mu\nu}(n)+
(-1)^2 \sum_{n_2=1}^M \sum_{n_1=1}^{n_2-1}
\phi_{\mu\lambda}(n_2) \phi_{\lambda\nu}(n_1) + \ldots \;.
\end{equation}

\vspace{0.5cm}
{\large \bf 3. Theoretical considerations surrounding the computation}

The object of computation is the quantity $F^{(M)}$, as given by eq.~(2).
The main effort amounts to carrying out the Grassmann integrations entering
this expression.
In between there intervenes the task of performing trace
operations over Lorentz indices associated with the loop(s). The tracing
involves strings of the $\phi_{\mu\nu}(n)$, cf.~eq.~(6).
The Lorentz trace can be postponed until after the Grassmann integration. This
trace accounts for setting the first and the last Lorentz index in a series
of products of the objects $\phi_{\mu\nu}(n)$ equal.
Since the second index of a $\phi_{\mu\nu}(n)$ factor must be saturated with
first index of the factor that follows, the Grassmann integrations will be
carried out first if the specific order by which the sequence of
the $\phi_{\mu\nu}(n)$'s are placed in each product is kept undisturbed.
Accordingly, the $\phi_{\mu\nu}(n)$ are considered non-commutative objects
during the Grassmann integration. By activating this rule the Lorentz indices
can be dropped until after the Grassmann integration has been completed.

To get a concrete handle on the situation, we introduce the quantities
\begin{equation}
A_{n}=-2(\varepsilon_\mu^n p_{n,\nu}-\varepsilon_\nu^n p_{n,\mu})\;,
\end{equation}
\begin{equation}
B_{n}=-\frac{4}{T} (\varepsilon_\mu^{n+1} \varepsilon_\nu^n -
\varepsilon_\nu^{n+1} \varepsilon_\mu^n ) \delta(u_{n+1}-u_n)\;,
\end{equation}
\begin{equation}
C_{n}=\sum_{m \neq n} \varepsilon^n \cdot p_m \partial_n G(u_n,u_m)
\end{equation}
and
\begin{equation}
D_{nm}=\frac{1}{2T}
\varepsilon^n \cdot \varepsilon^m \partial_n \partial_m G(u_n,u_m)\;.
\end{equation}
Using the fact that for commutative objects $\exp (C+D)=\exp(C) \exp(D)$,
the Grassmann calculation assumes the form

\[
F^{(M)}=\left[\prod_{n=M}^1 \int d\xi_n d\bar{\xi}_n\right]
\left\{ P\exp\left[\sum_n (
\bar{\xi}_n\xi_n A_{n}
+\bar{\xi}_{n+1}\xi_{n+1} \bar{\xi}_n\xi_n B_{n}
) \right] -2\right\}
\]
\begin{equation}
\exp\left[\sum_{n} \bar{\xi}_n \xi_n C_{n} \right]
\exp\left[\sum_n \sum_{m \neq n} \bar{\xi}_n \xi_n \bar{\xi}_m \xi_m D_{nm}\right]\;.
\end{equation}

A direct way to proceed with this calculation is to suitably code the
functions and the Grassmann variables and then separately calculate the
objects
$\exp\left[\sum_{n} \bar{\xi}_n \xi_n C_{n} \right]$
\newline$\exp\left[\sum_n \sum_{m \neq n} \bar{\xi}_n \xi_n \bar{\xi}_m \xi_m D_{nm}\right]$
and
$\left\{ P\exp\left[\sum_n (\bar{\xi}_n\xi_n A_{n}
+\bar{\xi}_{n+1}\xi_{n+1} \bar{\xi}_n\xi_n B_{n}) \right] -2\right\}$.
Each of these objects may contain sums of products with at most $M$
Grassmann variables $\xi_n$.
Then the multiplication between them can be carried out.
Since the following Grassmann properties are valid
\begin{equation}
\int d\xi_n \xi_n=1,\;\int d\bar{\xi}_n \bar{\xi}_n=1,\;
\int d\xi_n=\int d\bar{\xi}_n=0,\;
\xi_n\xi_m=-\xi_m\xi_n,\;\bar{\xi}_n\bar{\xi}_m=-\bar{\xi}_m\bar{\xi}_n,
\end{equation}
the output of the integration becomes obvious. Only those terms survive which
contain exactly $M$ products of $\xi_n\bar{\xi}_n$ with all the indices
different from each other.
This calculation is straightforward and we have written out the
corresponding FORTRAN code for carrying it out [13]. The
routine is effective (we present in Fig.~3 results up to $M=10$), but the
time consumed to reach the result rises exponentially with larger values of $M$.
The basic obstacle to the effectiveness of the routine is that the plethora
of terms involved in the intermediate computations does not survive in the
final result. A second source of trouble is the consumed amount of time for
rearranging terms at each stage of the computation before identical ones are
gathered and summed.

In this paper the aforementioned issues are confronted by following a
different computational procedure, which is based on the following
objectives: (a) at each stage of the calculation only the surviving terms are
computed and (b) similar terms are grouped together in the final result.
The above goals become more concrete by making the following
observations concerning the form of the terms entering our final expressions.
These terms will, in general, be products of the quantities
$A_n$, $B_n$, $C_n$ and $D_{nm}$ accompanied by Grassmann variables the
origin of each one of which becomes evident from eq.~(14).
It is also apparent that each of
$A_n$ or $C_n$ is accompanied by one pair of Grassmann variables, whereas
two such pairs accompany $B_n$ and $D_{nm}$. Let $N_1$, $N_2$, $N_3$ and
$N_4$ be the number of the functions $A_n$, $B_n$, $C_n$ and $D_{nm}$,
respectively appearing in a given term of the final result.
According to the Grassmann properties, the terms entering this result
must satisfy
\begin{equation}
N_1+2N_2+N_3+2N_4=M\;,\;0 \leq N_i \leq M,\; N_3+N_4 \neq 1,\;
N_i\in {\bf N},\;i=1,2,3,4\;,
\end{equation}
with the condition $N_3+N_4 \neq 1$ coming from the fact that the Lorentz
trace of a single function $A_n$ or $B_n$ is zero.

The above equation is very important because it helps one to determine all
the possible forms of the final terms before actually carrying out the
calculation.
Every group of the integers $N_i$ represents a particular form of the
final terms, each of which will be referred to as ``structure''.
Finding all the possible groups of $N_i$ suffices to determine all the
possible structures. In this way the calculation is broken into a number
of subcalculations in each of which one collects all the terms that
correspond to a specific structure. A term belonging to a particular
structure cannot be similar to a term belonging to another structure, since
such terms will contain different numbers of functions $A_n$, $B_n$, $C_n$
and $D_{nm}$, so goal (b) is satisfied.

Let us turn now to goal (a). Since in the final result there must exist
exactly $M$ pairs of Grassmann variables, all different from each other,
it is evident that every part of one exponential entering eq.~(14) has to
be multiplied by those components of the rest of the exponentials which
satisfy eq.~(16). We shall try now to trace how many terms survive in a
specific structure characterised by the integers $N_i$.

Firstly, as far as the $P\exp$ function entering (14) is concerned,
one may ask how can one obtain exactly $N_1$ functions $A_n$ and $N_2$ functions
$B_n$ placed in an ordered product according to a specific time order.
It is better to count the terms of the substructure $N_1,N_2$ first because
the fact that the associated product is ordered imposes more restrictive
conditions. For every group of Grassmann pairs taken from the total $M$ pairs
which enter the substructure $N_1,N_2$, the remaining Grassmann pairs will
always form a substructure $N_3,N_4$ (according to (16)). The converse,
however, is not true: If a substructure $N_3,N_4$ is formed first, then the
remaining Grassmann pairs cannot always form a substructure with given $N_2$.
The reason is that the quantities $B_n$ are associated with the Grassmann factor
$\bar{\xi}_{n+1}\xi_{n+1} \bar{\xi}_n\xi_n$, connecting adjacent Grassmann
pairs. Consider, for example, the case where $N_2=1$ and the time order is of
the form 1, 2, 3, $\cdots$. If, now, we are left with the Grassmann indices
1, 3, after considering the previous terms, then it is obvious that such a
case will not lead to surviving terms.

Generally speaking there are
$\left( \begin{array}{c} N_1+N_2 \\ N_1 \end{array} \right)$
ways one can have products of $N_1$ $A_n$ functions and $N_2$ $B_n$ functions,
if one is not concerned with the index $n$ they carry.
This result counts the ways the $N_1$ $A_n$ functions (considered
indistinguishable) can occupy $N_1+N_2$ total
vacant places. The rest are then filled with the functions $B_n$.
It can be checked that for everyone of these products of functions the ways
the indices $n$ can be distributed among the functions of the same kind is
the same. So let us consider a product where the first $N_2$ functions are
of the kind $B_n$ (contributing $2N_2$ Grassmann pairs) and the last $N_1$
functions are of the kind $A_n$ (contributing $N_1$ Grassmann pairs).
If the first $2N_2$ Grassmann pairs (according to the given time order) are
attributed to the $N_2$ functions $B_n$, then there are $M-2N_2$ Grassmann
pairs left for the $N_1$ functions $A_n$ and there are
$\left( \begin{array}{c} M-2N_2 \\ N_1 \end{array} \right)$ ways this can be
achieved.
Let us now suppose that the first $2N_2+1$ Grassmann pairs have been used up
through $N_2$ functions $B_n$. This means that a single Grassmann pair is
missing, hence one of the $B_n$ should begin by skipping the
Grassmann pair that precedes it.
There are $N_2$ ways this can happen and the result is that there are
now $M-2N_2-1$ Grassmann pairs to be distributed among $N_1$ places.
The whole scheme goes on this way. Each time one uses $2N_2$ from the first
$2N_2+i$ Grassmann pairs in the $N_2$ $B_n$ functions. The ways this can be
done are equal to the ways one can distribute $i$ indistinguishable objects
at $N_2$ distinguishable places. The result is
$\left( \begin{array}{c} i+N_2-1 \\ N_2-1 \end{array} \right)$ and has to be
multiplied by
$\left( \begin{array}{c} M-2N_2-i \\ N_1 \end{array} \right)$, which are the
ways the remaining Grassmann pairs can be distributed in the $N_1$ places.
Summing for all $i$ we have the result
\begin{equation}
\sum_{i=0}^{M-N_1-2N_2}
\left( \begin{array}{c} i+N_2-1 \\ N_2-1 \end{array} \right)
\left( \begin{array}{c} M-2N_2-i \\ N_1 \end{array} \right)\;.
\end{equation}
Eq.~(17) is true when $N_2\neq0$. When $N_2=0$ the corresponding result is
simply the ways we can distribute the $M$ Grassmann pairs in $N_1$ places,
that is
\begin{equation}
\left( \begin{array}{c} M \\ N_1 \end{array} \right)\;.
\end{equation}
Since the products coming from the $P\exp$ function are ordered they are all
unique, which means that each one of them registers directly, i.e. without any weight factor.

Our next step is to consider how do products of exactly
$N_3$ Grassmann pairs $\bar{\xi}_n\xi_n$ attributed to
$\exp\left[\sum_{n} \bar{\xi}_n \xi_n C_{n} \right]$ arise, if the unused
Grassmann pairs from the substructure $N_1,N_2$ are $\nu_3=M-N_1-2N_2$.
Obviously, the latter products originate from the component of the exponential
term
\begin{equation}
\frac{1}{N_3!}\prod_{j=1}^{N_3}
\left(\sum_{n_j=1}^{\nu_3} \bar{\xi}_{n_j} \xi_{n_j} A_{n_j}\right)\;.
\end{equation}
The sum in the above expression is extended over all the Grassmann pairs that
have not been used in the previous substructure, since the entrance of anyone
of the rest $M-\nu_3$ pairs in this sum will have null contribution to
the final result.
In short, given that the double appearance of a Grassmann variable in any product
causes it to vanish, we need to find the number of terms that survive. This number is
equal to the number of the ways the $\nu_3$ Grassmann pairs can be distributed
in $N_3$ places. Each place, here, corresponds to the position the particular
Grassmann pair will enter the product. Generally speaking $\nu_3$
Grassmann pairs are allowed to occupy the first position. For every
Grassmann in the 1st position, $\nu_3-1$ Grassmann pairs are allowed to
occupy the 2nd, etc. All the allowed
products are therefore the orders of $\nu_3$ per $N_3$, that is $(\nu_3)_{N_3}=$
$\nu_3(\nu_3-1)\cdots(\nu_3-N_3+1)$. But these terms are not all different.
Two Grassmann pairs, given that they contain two Grassmann variables, may
change their position without having any effect on their relative sign. So
the products of $N_3$ pairs are identical if they contain the same Grassmann
variables. We may then collect these identical products. Their number coincides
with the number of the permutations of $N_3$ objects to $N_3$ places, that is
$N_3!$. The number of different terms from the part of the exponential given
by (18) diminishes then to combinations of $\nu_3$ per $N_3$
\[
\frac{(\nu_3)_{N_3}}{N_3!}=\frac{\nu_3(\nu_3-1)\cdots(\nu_3-N_3+1)}{N_3!}=
\frac{\nu_3(\nu_3-1)\cdots(\nu_3-N_3+1)(\nu_3-N_3)!}{N_3!(\nu_3-N_3)!}=
\]
\begin{equation}
=\left( \begin{array}{c} \nu_3 \\ N_3 \end{array} \right)\;.
\end{equation}
The factor accompanying each of these terms will be then $N_3!$, divided by
the same factor entering (19) in the denominator. The result will be an
overall unit factor.

Next we ask how do products of exactly $2N_4$ Grassmann pairs
$\bar{\xi}_n\xi_n$ enter through
\newline $\exp\left[\sum_{n \neq m} \bar{\xi}_n \xi_n \bar{\xi}_m \xi_m D_{nm}\right]$,
if the unused Grassmann pairs from the previous substructures are
$\nu_4=M-N_1-2N_2-N_3$. These products originate from the part of the
exponential
\begin{equation}
\frac{1}{N_4!}\prod_{j=1}^{N_4}
\left(\sum_{n_j=1}^{\nu_4} \sum_{{m_j} \neq {n_j}}
\bar{\xi}_{n_j} \xi_{n_j} \bar{\xi}_{m_j} \xi_{m_j} D_{n_jm_j}\right)\;.
\end{equation}
The number of terms that survive is the number of the ways we can
distribute the $\nu_4$ Grassmann pairs to $2N_4$ positions, since every function
$D_{nm}$ provides two positions of Grassmann pairs. 
The number of allowed products are the orders of $\nu_4$ per $2N_4$, that is
$(\nu_4)_{2N_4}$. In order to isolate different terms we must divide by the
ways the $N_4$ functions $D_{nm}$ could appear in the product. There are $N_4!$
such ways. But since every $D_{nm}$ function is equal to a $D_{mn}$
function, we have to divide by a factor of 2 for every such function. We have
$N_4$ such functions, so the total factor by which we divide is
$2^{N_4}$. Summarising, the number of surviving, different terms is
\begin{equation}
\frac{1}{2^{N_4}N_4!}(\nu_4)_{(2N_4)}=
\frac{(2N_4)!}{2^{N_4}N_4!}\left( \begin{array}{c} \nu_4 \\ 2N_4 \end{array}
\right)\;.
\end{equation}
The factor accompanying each of these terms will then be $2^{N_4}N_4!$,
divided by the $N_4!$ which appears in (21), the net result being an overall
factor of $2^{N_4}$.

Collecting the above findings we arrive at the conclusion that the total
number of terms in a structure $N_1$, $N_2$, $N_3$, $N_4$ that survive, while
being different, is
\begin{equation}
R(M,N_1,N_2)
\;\left( \begin{array}{c} N_1+N_2 \\ N_1 \end{array} \right)
\left( \begin{array}{c} M-N_1-2N_2 \\ N_3 \end{array} \right)
\frac{(2N_4)!}{2^{N_4}N_4!}\left( \begin{array}{c} M-N_1-2N_2-N_3 \\ 2N_4
\end{array} \right)\;,
\end{equation}
where
\begin{equation}
R(M,N_1,N_2)=\left\{ \begin{array}{l}
{\displaystyle \sum_{i=0}^{M-N_1-2N_2}}
\left( \begin{array}{c} i+N_2-1 \\ N_2-1 \end{array} \right)
\left( \begin{array}{c} M-2N_2-i \\ N_1 \end{array} \right)
,\;{\rm if}\;N_2\neq0 \\
\left( \begin{array}{c} M \\ N_1 \end{array} \right) 
,\;{\rm if}\;N_2=0 \end{array} \right. \;.
\end{equation}
The accompanying factor for every term is $2^{N_4}$.

The above considerations are important in the building up of the program. First
the knowledge of the factor accompanying each term is necessary, since in
this way one avoids producing identical terms and subsequently summing them up.
Also the number of different terms existing in each substructure is needed
to parameterise the dimensions of the arrays that are used in the
communication between the subroutines of the program, as it will become
apparent in the next section. Moreover, the theoretical prediction of the
number of terms expected in the final result is used to verify the output of
the program. This is most important especially for large $M$, where there are
numerous surviving terms.

\vspace{0.5cm}
{\large \bf 4. Presenting the structure of the program }

The program has been written in FORTRAN 90. The reason for this choice is
the use of allocatable arrays in all the subroutines. This enables us to
parameterise the dimensions of these arrays for different values of $M$,
since the size of some of them grows drastically with $M$. We are also able
to manage the available memory effectively.

According to our previous analysis we shall first produce in the program all
the groups of integers $N_1$, $N_2$, $N_3$, $N_4$ that define via eq.~(16)
the compatible structures for given $M$. This is accomplished in the
subroutine STRUCTUREMATRIX, which performs four loops at which the integers
$N_1$ and $N_3$ acquire all values from $0$ to $M$ and the integers $N_2$
and $N_4$ all values from $0$ to $M/2$. Once a successful
combination for the values of these integers is found, according to (16), it
is copied to a temporary matrix, for which its column dimension is known a
priori. This dimension is the number of all the different structures that
exist for a particular $M$, a number that grows relatively slowly with
$M$.\footnote{For example it equals 486 for $M=20$.} This number can be easily
evaluated for all $M$ and has been inserted to the data of the program
for all $M\leq20$. For $M>20$ the program calls, where it is needed,
subroutine DIMOFSTRUCTUREMATRIX, which produces the number of different
structures for the given $M$.
Given that we are interested in grouping the resulting terms
according to their power of $T$, we group the structures
according to this power. It is easily checked from eqs.~(1), (11) and (13) that
this power is $M-3-N_2-N_4$. The final result is inserted into a matrix
which is the output of STRUCTUREMATRIX.

The surviving terms in the final result are produced in subroutine
MULTIPLYALL. The latter first calls subroutine STRUCTUREMATRIX
and then for every particular structure it calculates first the compatible
substructure $N_1,N_2$, followed by the compatible substructure $N_3$ and
finally the compatible substructure $N_4$.

Three complementary routines to MULTIPLYALL have been constructed. The
first is subroutine COMBINATIONASCEND, which produces all the combinations
of the $L$ natural numbers from 1 to $L$ in groups of $K$ and places them in
ascending order in the lines of a matrix which constitutes the output of the
subroutine. An array $A$ with $K$ elements is used for temporary inputs.
The numbers from 1 to $K$ are placed in the first line of the output matrix,
with the last $K$-th column being marked as the change column. This means that
the lines which follow in the output matrix are written with the first $K-1$
elements the same while the $K$-th element acquires all the values from $K+1$
to $L$. The routine is repeated with the change column lowered by one.
If the change column is the $J$-th, it is checked whether
\begin{equation}
A_{J}\leq M-N+J\;.
\end{equation}
If (25) holds, the elements with index greater than $J$ acquire the values
\begin{equation}
A_{J+I}=A_{J}+I,\;(I=J+1,K)\;.
\end{equation}
Then the array $A$ is placed in the output and $K$-th column is marked again as
the change column, repeating the procedure. In the opposite case, where
(25) is false, the change column is lowered by one, until condition (25) is
fulfilled. An example of the output of this subroutine is given in Table 1.

The second subroutine is ORDERS, which finds the orders of $L$ natural
numbers from 1 to $L$ in groups of $K$ and places them in the lines of the
output matrix $C$. The matrix $C$ is filled column by column. In every
column two integers determine the filling course: The first is the
number of identical elements of the previous column. This integer is
called STEPGREAT and for the $i$-th column it equals
$\frac{(L-i+1)!}{(L-K)!}$. This expression gives the number of the orders of
the $L-i+1$ numbers which are left after the first $i-1$ have correspondingly
occupied the first $i-1$ columns.
The second integer, called STEPSMALL, is the number of repetitions
of the same element in the $i$-th column and equals STEPGREAT divided by
$L-i+1$. An example of the gradual filling of matrix $C$ is shown in Table 2.

The third subroutine is COMBINATIONPERTWO which finds the combinations of
the $L'$ natural numbers from 1 to $L'$ in groups of $K'$ after they have been
grouped in couples. Each group is placed in ascending order in the output matrix
$C$. Each couple always enters as $nm$ with $n<m$. Matrix $C$ occupies $2K'$
columns so that each couple is located at adjacent columns.
To achieve this COMBINATIONPERTWO calls first COMBINATIONASCEND
(with $L=L'$ and $K=K'$) and stores the result in matrix $CA$.
Then it calls ORDERS (with $L=L'-K'$ and $K=K'$) and stores the result in
matrix $CB$. Matrix $CA$ produces the odd columns of matrix $C$, that is
the first elements of the couples. For every line of $CA$ the remaining $L'-K'$
elements are found and placed in a temporary array $B$. The second element of
every couple is found from matrix $CB$ after its elements are read as
indices of the array $B$. Because of this last action subroutine ORDERS is
only called once in COMBINATIONPERTWO.  In the production of the group of
couples it is checked whether the second element is greater than the first.
If this is not so the subroutine proceeds to the next combination.
An example of the output of the subroutine is given in Table 3.

With the three complementary subroutines that have been described above the
calculation of the substructures can be carried out. First the calculation
of the substructure $N_1,N_2$ is produced.
By calling COMBINATIONASCEND from  MULTIPLYALL the
combinations of the positions of the $N_2$ $B_n$ functions from a total of
$N_1+N_2$ vacancies are found and placed in the matrix CPL. For every such
combination, which is represented as a line of CPL, one has to produce
all the compatible products of functions $A_n$ and $B_n$.
First we shall assume that the integers $n$, which characterise the functions
$A_n$ and $B_n$, run from 1 to $M$ and that they must appear in an ascending
order, from small to large numbers. We then
allocate two matrices CTEMP1 and CTEMP2 with equal dimensions. The number of
columns equals $N_1+N_2$ and the number of lines is given by eq.~(17) or (18).
The last value is returned by function COUNTTERMS2.
For every element in the positions from 1 to $N_1+N_2$ the maximum and 
minimum allowed values are found and are placed in the arrays IMAX and IMIN
respectively. One has to bear in mind that every function $B_n$ consumes two
consequent integers, so when one such function, characterised by integer
$n$, appears in the $i$-th place, then the integer in the $(i+1)$-th place
cannot be smaller than $n+2$.
The calculation begins by inserting in the first column of matrix CTEMP1 and
in the different lines all the allowed values for the first integer, as they are
dictated by the arrays IMAX and IMIN. In the second run all the allowed
combinations of the first two elements will be copied in matrix CTEMP2.
To this end, for each line of CTEMP1 the allowed values of the next
element are determined. These allowed values can range from the maximum to the
minimum value of the arrays IMAX and IMIN but they must also fulfil
additional constraints which depend on the preceding integer. If the
preceding integer is $n$, then the next integer will be $n+1$ when
the corresponding function is of the type $A$ and $n+2$ if it is
of type $B$. So each line of CTEMP1 does not necessarily correspond
to the same number of lines of CTEMP2. The calculation proceeds
with the combinations of the three first elements copied back to CTEMP1
with the data taken from CTEMP2 and so forth. Once all the elements are
exhausted, their allowed combinations will have been accommodated in one of the
matrices CTEMP1 or CTEMP2 and are then copied in matrix C12.
In the following example we suppose that $M=5$, $N_1=2$, $N_2=1$ and that
the lone function $B$ occupies the middle place among the three available for
the functions $A$ and $B$. The array IMIN has, subsequently, acquired the
values $\left( \begin{array}{ccc} 1&2&4 \end{array} \right)$
and the array IMAX  the values
$\left( \begin{array}{ccc} 2&3&5 \end{array} \right)$.
The calculation of the allowed combinations proceeds as follows
\[
\stackrel{\rm CTEMP1}
{\left( \begin{array}{c} 1\\2 \end{array} \right)} \rightarrow
\stackrel{\rm CTEMP2}
{\left( \begin{array}{cc} 1&2\\1&3\\2&3 \end{array} \right)} \rightarrow
\stackrel{\rm CTEMP1}
{\left( \begin{array}{ccc} 1&2&4\\1&2&5\\1&3&5\\2&3&5 \end{array} \right)}\;.
\]

The compatible substructure $N_3$ for every line in the substructure
$N_1,N_2$ is produced with the call of the subroutine
COMBINATIONASCEND with $L=M-N_1-2N_2$ and $K=N_1$ and the output is
stored in matrix $C1$. Also, for every line of CPL the remaining
elements which have not been used are determined and placed in the array LEFT34.
Matrix $C1$ contains the indices of the array LEFT34, i.e. the
presence of 2 in a line of $C1$ has the meaning of the 2nd element of
array LEFT34. With this representation the subroutine
COMBINATIONASCEND is called only once from MULTIPLYALL to produce the
substructure $N_3$.

The compatible substructure $N_4$ for every line in the substructures
$N_1,N_2$ and $N_3$ is produced with the call of the subroutine
COMBINATIONPERTWO with $L'=M-N_1-2N_2-N_3$ and $K'=N_4$ and the output is
stored in matrix $C2$. Also for every line of CPL and every line of
C1 the remaining
elements which have not been used are found and placed in the array LEFT4.
Matrix $C2$ contains the indices of the array LEFT4, i.e. the
presence of 2 in a line of $C2$ has the meaning of the 2nd element of
array LEFT4. With this representation the subroutine
COMBINATIONPERTWO is called only once from MULTIPLYALL to produce the
substructure $N_4$.

Until now we have described the production of the sequence of functions
$A$, $B$, $C$ and $D$ in the terms that survive in the final result. The indices
$n$ which characterise the functions in this description are considered as
natural numbers from 1 to $M$ and their order is considered to be an
ascending one. Clearly, the order of these natural numbers is the time
order given a priori by the user. This time order is stored in the array
TOR. So in order to obtain the final result we have to read the elements in
the structure which has been described as indices of the array TOR.
The output of MULTIPLYALL is, for every structure, a series of lines which
represent the products of the functions (10)-(13). Their representation, which
is given by integers, is listed in Table 4. It should be noted that even
though the functions $A_n$, $B_n$ and $D_{nm}$ may happen to be represented by
the same integer, they cannot be confused, because they have
a specific order of appearance. Specifically the non-commutative objects of
the substructure $N_1,N_2$ appear first, followed by the commutative objects of
the structure $N_3$ and finally by those of the structure $N_4$.

The output of MULTIPLYALL is written in a series of files with the
first one designated as F001. When a file exceeds 18,000,000 terms then
the next structure will be written in the next file. Each line in a
structure is accompanied by a factor which is just $2^{N_4}$, according to
the discussion of the previous section. In the cases for which the substructure
$N_1=N_2=0$ enters, this factor is multiplied by 2. This comes from
$Tr_L (\delta_{\mu\nu})-2=4-2=2$. The output of MULTIPLYALL is written
in the form of integers and so it can be used for further calculations by the
user.

The program gives in the output of MULTIPLYALL the products of surviving
functions. However the Lorentz trace has yet to be performed. This
trace involves all the functions in the $N_1,N_2$ substructure and this is the
reason these objects have been considered as being non commutative.
The Lorentz indices $\mu$, $\nu$ in the functions (10), (11) have been
considered indefinite, that is they have not been defined at this point by
the number which represents the function. So when groups of these types of
functions are found together it is assumed that the Lorentz trace has to
be performed, according to the exact order by which the corresponding
functions appear.
The following example illustrates the ``translation'' of a typical product of
three such functions
\begin{equation}
202\;301\;204\;\rightarrow\;
(202)_{\mu\rho}\;(301)_{\rho\sigma}\;(204)_{\sigma\mu}\;\rightarrow\;
(A_2)_{\mu\rho}\;(B_1)_{\rho\sigma}\;(A_4)_{\sigma\mu}\;.
\end{equation}

In order to perform the Lorentz trace and arrive at the final output
we have constructed the subroutine TRACE.
For the computation of the trace only the part
$(\varepsilon_\mu^n p_{n,\nu}-\varepsilon_\nu^n p_{n,\mu})$ from functions
$A_{n}$ and the part 
$(\varepsilon_\mu^{n+1} \varepsilon_\nu^n -
\varepsilon_\nu^{n+1} \varepsilon_\mu^n )$ from functions $B_{n}$ need to be
considered.
The following representation is used
\begin{equation}
\varepsilon_\nu^n\;\rightarrow \;300+n\;,\;\;\; p_{n,\nu}\rightarrow \;200+n\;,
\end{equation}
where $n$ is a two digit integer. Then the relevant part of the functions
$A_{n}$ and $B_{n}$ for the trace is represented by two matrices
\begin{equation}
\varepsilon_\mu^n p_{n,\nu}-\varepsilon_\nu^n p_{n,\mu} \rightarrow
\left(\begin{array}{rr}1\\-1\end{array}\right)
\left(\begin{array}{rr}300+n&200+n\\200+n&300+n\end{array}\right)\;,
\end{equation}
\begin{equation}
\varepsilon_\mu^{n+1} \varepsilon_\nu^n -
\varepsilon_\nu^{n+1} \varepsilon_\mu^n 
\rightarrow
\left(\begin{array}{rr}1\\-1\end{array}\right)
\left(\begin{array}{cc}300+n+1&300+n\\300+n&300+n+1\end{array}\right)\;.
\end{equation}
In a product of functions $A_{n}$ and $B_{n}$, where the Lorenz indices are
active, the second index of one function is saturated with the first index of
the next. This continues until the last function in the product is reached,
whereupon its second index is saturated by the first index of the first
function.
To accomplish this in the representation (29)-(30) it suffices to perform
internal products between the elements of the 2nd column of the 2x2 matrix that
represents one function and the 1st column of the 2x2 matrix representing the
next function. The elements of the 2nd column of the last matrix are coupled
with the elements of the 1st column of the first. The procedure has been
appropriately parameterised in the program to exhaust all the combinations.

The rule $\varepsilon^n \cdot p_n = 0$ is also taken into account. On the
other hand, we have not applied the replacement associated with momentum
conservation, since it spoils the symmetry of the terms belonging to the
same structure in addition to increasing the number of terms, especially for large $M$.

The result of the trace, for each term, corresponds to the sums of products of internal
products of $\varepsilon$'s and $p$'s. These sums are again represented by
series of integers. An internal product is represented by a six digit integer,
according to the prescription of Table 5.

The factor of a product is represented by an integer less than or equal to
200000. Such factors also signal both the beginning of the product of
internal products, as well as, its ending. This can be summarised in the next
example
\[
\begin{array}{cccccccccc}
-1&304301&1&304303&302203&301202&204204&-3&304203&303302 \end{array} \; \rightarrow
\]
\begin{equation}
-\varepsilon^4\cdot\varepsilon^1+
\varepsilon^4\cdot \varepsilon^3\;\;\varepsilon^2\cdot p_3
\;\;\varepsilon^1\cdot p_2\;\;p_4\cdot p_4
-3\;\;\varepsilon^4\cdot p_3\;\;\varepsilon^3\cdot \varepsilon^2\;.
\end{equation}

The powers of 2 and $T$ associated with the functions (10), (11) and (12) are
incorporated into the factor of every term. The representation of the
remaining functions are given through integers according to Table 6.

The output of TRACE is registered in two series of files. The first
series begins with file G001 and it contains the result of surviving terms
grouped according to the structure they belong and are represented through
integers according to Tables 5 and 6. The content of these files, as it is in
integer form, may be used for further processing by the user.
The second series begins with file H001 and
it contains the surviving terms in function-like form. The terms entering
files G001 and H001 correspond to the terms stored in files F001,
and so on. There is, also, the option to produce with TRACE only the
first term of each structure. In that case TRACE avoids reading through
the whole series of terms produced by MULTIPLYALL. Instead it calls
MULTIPLYALL again and asks for the production of only the first
term of each structure. MULTIPLYALL writes the result in only one file,
FF001, which is then read by TRACE. With this procedure the 
timesaving in TRACE is enormous.

The different structures that exist for the current $M$ are listed in file
STRLOG, along with information about the files used for the output and the
consumed time for the run of the program.

\vspace{0.5cm}
{\large \bf 5. Results}

At the beginning of its run the program asks the user to insert the
number of external gluons $M$.
Because the output of subroutine MULTIPLYALL occupies storage space
which increases rapidly with $M$, the user is warned about the approximated
space that will be required. This space has been measured from the output
of the program until some value of $M$. For greater values of $M$ it is
extrapolated through a fitting function of the form
$y=\exp [a(\ln(x))^2+b\,\ln(x)+c]$, where $y$ is the storage space, $x$ the
total number of surviving terms and $a$, $b$ and $c$ free parameters which
have been determined through a fit on the existing data. The same function
has been used for the extrapolation on the storage space required by
TRACE. Then the program requests for the specific
time order for which the calculation will be carried out and a check is
performed on the correct input of this time order.

Because the output of subroutine TRACE occupies large storage space
the user has the option to ask for the full result of this subroutine, or for
the calculation of only the first term of every structure, or to skip the
subroutine entirely. Each choice of output of TRACE is accompanied by
a warning about the required storage space.

In Table 7a we show the result of MULTIPLYALL for $M=2$ and in Table 7b
the result of TRACE in function-like form for the same value of $M$.
The 23 terms for $M=3$ are shown in function-like form in Table 8.
The number of surviving terms grows rapidly when $M$ increases
(for $M=15$ the surviving terms are 430,576,126).

In Figure 2 we have plotted
the number of surviving terms (left axis - curve (a)) against $M$. In the
same figure we have plotted the consumed storage space for the output of
MULTIPLYALL (right axis - curve (b)), which is more compact than the full
final result of TRACE.
At $M=15$ this space grows to about 22.2 Gbytes, making difficult for us to
have the full result written on our hard disc for greater $M$'s.

In Fig.~3 we have plotted the number of structures as a function of $M$ (left
axis - curve (a)). On the same graph we have depicted the average number of surviving
terms per structure as a function of $M$ (right axis - curve (b)). It is seen that the
number of different structures does not grow rapidly, in contrast to the
surviving terms per structure. So the time needed for the completion of the
program is mainly due to the intense growth of number of these terms.

In Fig.~4 we have plotted the consumed time for the completion of the
calculation of MULTIPLYALL (full line) against $M$ (curve (a)).
The consumed time is the most crucial parameter since it
exhibits the capabilities of our program, especially for large $M$.
This is only $2.74$ seconds for $M=10$ and $1.75$ hours for $M=15$ (on INTEL
CELERON 1.8GHz).
The capability of the current program is more evident if it is compared with
the older program we had developed in [13] and where the calculation is carried
out in the straightforward way without the use of structures. With respect to
that program we have made the substitution of eq.~(12) in order to group
functions with the same Grassmann
part in the same overall function, as the case is for the present program and
have recorded the time consumed for the calculation without the subroutine
TRACE. This allows a comparison to be carried out on the same
footing. The result is the dotted line (curve (b)). The slashed line in
Fig.~4 (curve (c)) is the consumed time in our current program for
MULTIPLYALL to be completed but with suppressed writing in the
output files which occupy enormous storage space. In this way we carry
calculations for greater values of $M$. It is also evident, from the
comparison of curves (a) and (c) that a large
fraction of the consumed time for MULTIPLYALL is due to the output.
In curves (a)-(c) we have used only the time consumed in MULTIPLYALL and
have not added the time consumed in TRACE because the larger part of the
time in the latter is consumed in writing the output in the appropriate
files and not in the actual calculation. The consumed time in TRACE as
function of $M$ is presented by curve (i) for the full result and by curve
(ii) for the result with only one term per structure.

In Fig.~5 we have plotted the consumed time, as in Fig.~4, but against the
surviving terms. The consumed time (full line) in our new program as a function
of the surviving terms can be fitted through a linear function.
The consumed time in our old program grows much more rapidly as function
of the surviving terms (dotted line). In that case it can be fitted by a
second grade polynomial which has been extrapolated to larger values of $M$.
This fitting procedure leads to a value of about $1910$ years for $M=15$
and can be compared to the 1.75 hours for the same result with our new
program. The difference is due to the appearance in the intermediate
calculations of our old program of a vast number of terms which cancel in the
final result.

In the overall result divergent terms enter only for $M\leq4$. These
terms have been compared with the analytic calculations for the
divergent terms for $M=2$, $M=3$ and $M=4$ [10] and have been found in total
agreement. Also the number of surviving terms according to the program have
been compared with the theoretical formulas (23)-(24) and have been found
equal. The number of surviving terms between the old and the new program have
also been found in agreement.

The methods used for the computation of the master formulas do not place an
upper bound on the value of $M$ for the calculations. In the specific
program we have written the different representations through integers
assuming that $M$ is a two digit integer.
The limitations are imposed mainly by the storage space and of course,
unavoidably, by the necessary time which, even for linear growth, will reach
high values for increasing $M$'s. The consumed time, though, is expected to shrink
proportionally to the increase of the speed of the processor to be used.
Another limitation arises from the maximum integer that can be encoded in the
processor one uses for the running of the program. This sets an upper
bound on the number of the elements that can be allocated to a certain array.
The size of the matrix that grows more rapidly is the output of
COMBINATIONPERTWO which is called from MULTIPLYALL to produce the
substructure $N_4$. In processors that use 4-byte integers overflow is
expected to arise for $M\geq20$, while in processors that use 8-byte integers
the overflow is expected to appear for $M\geq34$.
It is possible to improve these values of $M$, in any case, by defining the
respective array in a more complicated way, but we have avoided
doing so, since the considered values of $M$ are already quite high.

\vspace{0.5cm}
{\large \bf 6. Concluding remarks}

In this paper a computational algorithm has been presented aiming towards
calculation of one-gluon loop Feynman diagrams, including
ghost loop contributions, with $M$ external gluon attachments, on the basis
of the master formulas derived in [10]. Compared with our previous attempt
[13], the reasoning of the calculation has been completely changed, resulting to a
different algorithm. The basic innovations are the grouping of similar terms
together and the dramatic reduction of computation time, thus allowing the
calculation for quite large values of $M$.
All this being said, there remains, of course, the problem of carrying out the
integrations over the Feynman parameters, a task which is challenging in itself, 
both on the analytical and the numerical front. With respect to the latter, it
would be of interest to assess the extent to which recent, relevant, 
considerations [14] can be applied for the succesful completion 
of the task, at least up to some reasonably high value of $M$, 
whose Grassmann variable integration part was accomplished in this work.

Finally, one may express the hope that the particular feature of the constructed
algorithm, namely the ability to expedite integrations over a multivariable
set, a subset of which is Grassmannian, could find wider applications to
analogous situations that may arise in other physical problems wherein
Grassmann variables make their entrance. Within the context of the present
application, it would be of interest to apply the particular algorithm
developed in this work to the two-gluon loop $M=4$ case, the corresponding
master expressions for which have been derived in [11]. As a first
attempt, one could restrict the relevant computation to the divergent term
associated with the $M=2$ configuration and verify the consistency with
second order corrections to the running coupling constant in pQCD.

\vspace{0.3cm}
{\large \bf Acknowledgement}

A.~I.~K.~and C.~N.~K.~acknowledge the financial support through the program
``Pythagoras I'' (grant no.~016) and from the General
Secretariat of Research and Technology of the University of Athens.

\newpage
{\large \bf Tables}

\[
\left( \begin{array}{ccc} 1&2&3\\1&2&4\\1&3&4\\2&3&4 \end{array} \right)
\]
\begin{center}
{\it Table 1. The output matrix of subroutine COMBINATIONASCEND with
$L=4$ and $K=3$.}
\end {center}

\vspace{0.3cm}
\[
\left( \begin{array}{cc} 1&0\\1&0\\2&0\\2&0\\3&0\\3&0 \end{array} \right)
\rightarrow
\left( \begin{array}{cc} 1&2\\1&3\\2&1\\2&3\\3&1\\3&2 \end{array} \right)
\]
\begin{center}
{\it Table 2. The gradual filling of the output matrix of subroutine
ORDERS with $L=3$ and $K=2$.}
\end {center}

\vspace{0.3cm}
\[
\left( \begin{array}{cccc} 1&3&2&4\\1&4&2&3\\1&2&3&4 \end{array} \right)
\]
\begin{center}
{\it Table 3. The output matrix of subroutine COMBINATIONPERTWO with
$M=4$ and $N=2$.}
\end {center}

\begin{center}
\begin{tabular}{cr} \hline
Function & Representation \\ \hline
$A_n$    & $200+n\;\;\;\;$    \\ 
$B_n$    & $300+n\;\;\;\;$    \\ 
$C_n$    & $n\;\;\;\;$        \\ 
$D_{nm}$ & $100*n+m\;\;\;\;$  \\ \hline
\end{tabular} 
\end{center} 

\begin{center}
{\it Table 4. The representation of functions through integers in the output
of subroutine MULTIPLYALL. The numbers $n$, $m$ are to be considered as
two digit integers.}
\end {center}

\newpage
\vspace{0.5cm}
\begin{center}
\begin{tabular}{cc} \hline
Internal                            & Representation    \\
Product                             &                   \\ \hline
$p_n \cdot p_m$                     & 200000+n*100+200+m \\
$\varepsilon^n \cdot p_m$           & 300000+n*100+200+m \\
$\varepsilon^n \cdot \varepsilon^m$ & 300000+n*100+300+m \\ \hline
\end{tabular} 
\end{center} 

\begin{center}
{\it Table 5. The representation of internal products through integers
in the output of TRACE.}
\end {center}

\vspace{0.5cm}
\begin{center}
\begin{tabular}{cc} \hline
Function                                                                           & Representation \\ \hline
$\displaystyle \delta(u^{n+1}-u^n)$                                                & 300+n            \\
$\displaystyle \sum_{m \neq n} \varepsilon^n \cdot p_m \partial_n G(u_n,u_m)$      & 20000+n*100+m    \\
$\displaystyle \varepsilon^n \cdot \varepsilon^m \partial_n \partial_m G(u_n,u_m)$ & 30000+n*100+m    \\
                                                                                   & (=30000+m*100+n) \\ \hline
\end{tabular} 
\end{center} 

\begin{center}
{\it Table 6. The representation of functions through integers in the output
of TRACE.}
\end {center}

\vspace{0.5cm}
\renewcommand{\baselinestretch}{1.2}
{\scriptsize

\begin{tabular}{ll}                                     \hline
 M= 2,     TIME ORDER: 1 2          &                 \\
 STRUCTURE:( 0  0  0  1) [T**( -2)] &                 \\
 TERMS=                             &1.               \\
 FACTOR=                            &\hspace{0.1cm} 4.\\
 \hspace{0.1cm}102                  &                 \\
 STRUCTURE:( 0  0  2  0) [T**( -1)] &                 \\
 TERMS=                             &1.               \\
 FACTOR=                            &\hspace{0.1cm} 2.\\
 \hspace{0.4cm}1 \hspace{0.4cm}2    &                 \\
 STRUCTURE:( 2  0  0  0) [T**( -1)] &                 \\
 TERMS=                             &1.               \\
 FACTOR=                            &\hspace{0.1cm} 1.\\
 \hspace{0.1cm}202 \hspace{0.2cm}201&                 \\
 TOTAL \#  OF                       &                 \\
 TERMS=                             &3.               \\ \hline
\end{tabular}

}

\begin{center}
{\it Table 7a. The output of MULTIPLYALL for $M=2$ and the time order
$u_1<u_2$.}
\end {center}

\newpage
\renewcommand{\baselinestretch}{1.2}
{\scriptsize

\vspace{1cm}
\hspace{0.1cm}M= 2, TIME ORDER: 1  2

\underline{\hspace{8cm}}

\hspace{0.1cm}-(pi**2/2)*g**(2 )*d4(p1 +p2 )*TrC(tGa2  tGa1  )*

\begin{tabular}{llllrccl}
 inf       & 1          & u2         &             &         & 2 &  2  &                   \\
 $\mid$ dT & $\mid$ du2 & $\mid$ du1 &  8(u2 ,u1 ) &* Exp\{T & S &  S  & [pn.pmG(un,um)]\}* \\
 0         & 0          & 0          &             &         &n=1&m=n+1&                   \\
\end{tabular}

\vspace{0.1cm}
\hspace{0.1cm} STRUCTURE:( 0  0  0  1) TERMS= \hspace{3.8cm} 1.

\hspace{5cm} 2. \hspace{0.2cm} T**(-2 )

\hspace{0.1cm} \{+ e1 .e2 d1 d2 G(u1 ,u2 ) \hspace{1cm} \}

\hspace{0.1cm} STRUCTURE:( 0  0  2  0) TERMS= \hspace{3.8cm} 1. 

\hspace{5cm} 2. \hspace{0.2cm} T**(-1 )

\hspace{0.1cm} \{+ S\_(i.ne.1 )[e1 .pid1 G(u1 ,ui)] \hspace{0.1cm} S\_(i.ne.2 )[e2 .pid2 G(u2 ,ui)] \hspace{0.1cm} \}

\hspace{0.1cm} STRUCTURE:( 2  0  0  0) TERMS= \hspace{3.8cm}  1. 

\hspace{5cm} 4. \hspace{0.2cm} T**(-1 )

\begin{tabular}{llllllllll}
 \{+&  &      &         &         &     &         &        &   &    \\
    &( &  +2  &  e2 .p1 & e1 .p2  &  -2 &  e2 .e1 & p2 .p1 & ) & \} \\
\end{tabular}

\underline{\hspace{8cm}}

\hspace{1.7cm} TOTAL \# of Terms = \hspace{3.8cm}  3.

}
\renewcommand{\baselinestretch}{1.5}

\begin{center}
{\it Table 7b. The function like output of TRACE for $M=2$ and the
time order $u_1<u_2$. The Memorandum is similar to Table 8.}
\end {center}

\newpage
\vspace{0.5cm}
\renewcommand{\baselinestretch}{1.2}
{\scriptsize

\hspace{0.1cm}M= 3, TIME ORDER: 3  1  2

\underline{\hspace{8cm}}

\hspace{0.1cm}-(pi**2/2)*g**(3 )*d4(p1 +p2 +p3 )*TrC(tGa3  tGa2  tGa1  )*

\begin{tabular}{lllllrccl}
 inf       & 1          & u2         & u1         &                 &         & 3 &  3  &                    \\
 $\mid$ dT & $\mid$ du2 & $\mid$ du1 & $\mid$ du3 &  8(u2 ,u1 ,u3 ) &* Exp\{T & S &  S  & [pn.pmG(un,um)]\}* \\
 0         & 0          & 0          & 0          &                 &         &n=1&m=n+1&                    \\
\end{tabular}

\vspace{0.1cm}
\hspace{0.1cm} STRUCTURE:( 0  0  1  1)  TERMS= \hspace{3.8cm}  3. 

\hspace{5cm} 2. \hspace{0.2cm} T**(-1 )

\hspace{0.1cm}\{+ S\_(i.ne.3 )[e3 .pid3 G(u3 ,ui)] \hspace{0.1cm}  e1 .e2 d1 d2 G(u1 ,u2 )        

\hspace{0.15cm}  + S\_(i.ne.1 )[e1 .pid1 G(u1 ,ui)] \hspace{0.1cm}  e3 .e2 d3 d2 G(u3 ,u2 )

\hspace{0.15cm}  + S\_(i.ne.2 )[e2 .pid2 G(u2 ,ui)] \hspace{0.1cm}  e3 .e1 d3 d1 G(u3 ,u1 ) \hspace{0.1cm} \}

\hspace{0.1cm} STRUCTURE:( 1  1  0  0)  TERMS= \hspace{3.8cm}  2.

\hspace{5cm} 8. \hspace{0.2cm} T**(-1 )

\hspace{0.1cm} \{+ d(u1 -u3 )

\begin{tabular}{llllllllll}
  & (   +2     &e3 .e2 & e1 .p2 &  -2  &  e3 .p2 & e2 .e1 &  )&   & \\
\end{tabular}

\hspace{0.25cm} + d(u2 -u1 )

\begin{tabular}{llllllllll}
  & (   +2     &e3 .e1 & e2 .p3 &  -2  &  e3 .e2 & e1 .p3 &  )&  \}& \\
\end{tabular}

\hspace{0.1cm} STRUCTURE:( 0  0  3  0)  TERMS= \hspace{3.8cm}  1.

\hspace{5cm} 2. \hspace{0.2cm} T**( 0 )

\hspace{0.2cm}\{+ S\_(i.ne.3 )[e3 .pid3 G(u3 ,ui)] \hspace{0.1cm} S\_(i.ne.1 )[e1 .pid1 G(u1 ,ui)] \hspace{0.1cm} S\_(i.ne.2 )[e2 .pid2 G(u2 ,ui)]\hspace{0.1cm} \}

\hspace{0.1cm} STRUCTURE:( 2  0  1  0)  TERMS= \hspace{3.8cm}  3.

\hspace{5cm} 4. \hspace{0.2cm} T**( 0 )

\hspace{0.2cm}\{+ S\_(i.ne.2 )[e2 .pid2 G(u2 ,ui)]    

\begin{tabular}{lllllllll}
  &  ( &  +2  &  e3 .p1 & e1 .p3  &  -2  &  e3 .e1 & p3 .p1 & ) \\
\end{tabular}

\hspace{0.35cm}+ S\_(i.ne.1 )[e1 .pid1 G(u1 ,ui)]

\begin{tabular}{lllllllll}
  &  ( &  +2  &  e3 .p2 & e2 .p3  &  -2  &  e3 .e2 & p3 .p2 & ) \\
\end{tabular}

\hspace{0.35cm}+ S\_(i.ne.3 )[e3 .pid3 G(u3 ,ui)]

\begin{tabular}{llllllllll}
  &  ( &  +2  &  e2 .p1 & e1 .p2  &  -2  &  e2 .e1 & p2 .p1 & )  &  \} \\
\end{tabular}

\hspace{0.1cm} STRUCTURE:( 3  0  0  0)  TERMS= \hspace{3.8cm}  1.

\hspace{4.9cm} -8. \hspace{0.2cm} T**( 0 )

\hspace{0.1cm}\{+  

\begin{tabular}{llllllllllllll}
  &  ( &  +1  &  e3 .p1 & e2 .p3 & e1 .p2  &  -1  &  e3 .e2 & e1 .p2 & p3 .p1  &  -1  &  e3 .e1 & e2 .p3 & p2 .p1 \\
  &    &  +1  &  e3 .e2 & e1 .p3 & p2 .p1  &  -1  &  e3 .p1 & e2 .e1 & p3 .p2  &  +1  &  e3 .p2 & e2 .e1 & p3 .p1 \\
  &    &  +1  &  e3 .e1 & e2 .p1 & p3 .p2  &  -1  &  e3 .p2 & e2 .p1 & e1 .p3  &)     &   \}    &        &        \\
\end{tabular}

\underline{\hspace{8cm}}

\hspace{1.7cm} TOTAL \# of Terms = \hspace{3.7cm}  10.

\underline{\hspace{16cm}}

Memorandum:

\hspace{-0.5cm}\begin{tabular}{cc|cc|cc|cc}
d4(p1+p2+p3 )                                                              &
$\delta^{(4)}(p_1+p_2+p_3)$                                                &
pi**2                                                                      &
$\pi^2$                                                                    &
d(u2 -u1 )                                                                 &
$\delta(u_2 -u_1 )$                                                        &
e2.e3                                                                      &
$\varepsilon^2 \cdot \varepsilon^3$                                        \\
TrC(tGa3 tGa2 tGa1 )                                                       &
$Tr_C (t_G^{\alpha_3} t_G^{\alpha_2} t_G^{\alpha_1})$                      &
inf                                                                        &
$\infty$                                                                   &
d2d3G(u2,u3)                                                               &
$\partial_2 \partial_3 G(u_2,u_3)$                                         &
e1.p2                                                                      &
$\varepsilon^1 \cdot p_2$                                                  \\
8(u2 ,u1 ,u3 )                                                             &
$\theta(u_2 ,u_1 ,u_3 )$                                                   &
S                                                                          &
$\sum$                                                                     &
T**(-1)                                                                    &
$T^{-1}$                                                                   &
p1.p2                                                                      &
$p_1 \cdot p_2$                                                            \\
S\_(i.ne.2 )[e2 .pid2 G(u2 ,ui)                                            &
$\sum_{i\neq 2}\varepsilon^2 \cdot p_i\partial_2 G(u_2,u_i)$               &
$\mid$                                                                     &
$\int$                                                                     &
                                                                           &
                                                                           &
                                                                           &
                                                                           \\
\end{tabular}

}
\renewcommand{\baselinestretch}{1.5}

\begin{center}
{\it Table 8. The function like output of TRACE for $M=3$ and the
time order $u_3<u_1<u_2$.}
\end{center}

\newpage
{\large{\bf Figures}}

\vspace{2cm}
\begin{center}
\includegraphics[scale=1,angle=0]{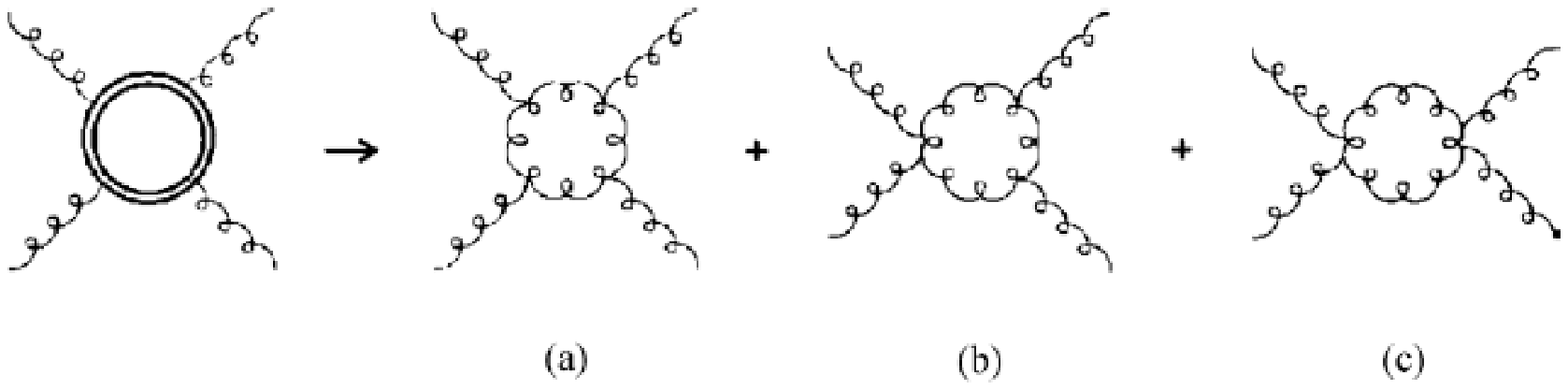}

\rm {\bf Figure 1} Illustration, for $M=4$, of the classes of one-gluon loop
Feynman diagram (right side of arrow) which are simultaneously accommodated by
the corresponding master formula depicted on left side of arrow.
\end{center}

\newpage
\begin{center}
\includegraphics[scale=1,angle=0]{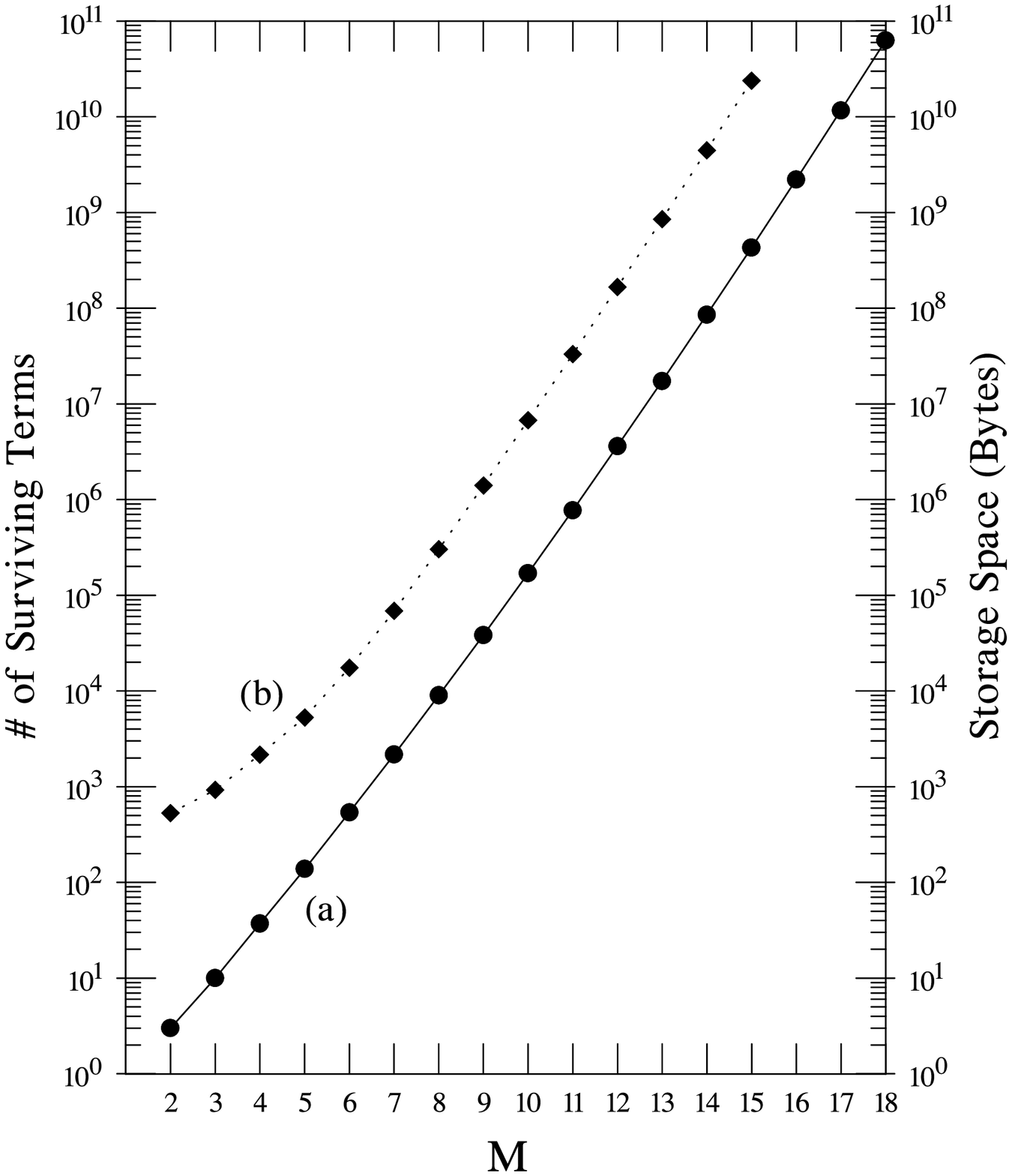}

\rm {\bf Figure 2} (a) The number of surviving terms (left axis) as function
of $M$. (b) The storage space occupied by the output of subroutine
MULTIPLYALL (right axis).
\end{center}

\newpage
\begin{center}
\includegraphics[scale=1,angle=0]{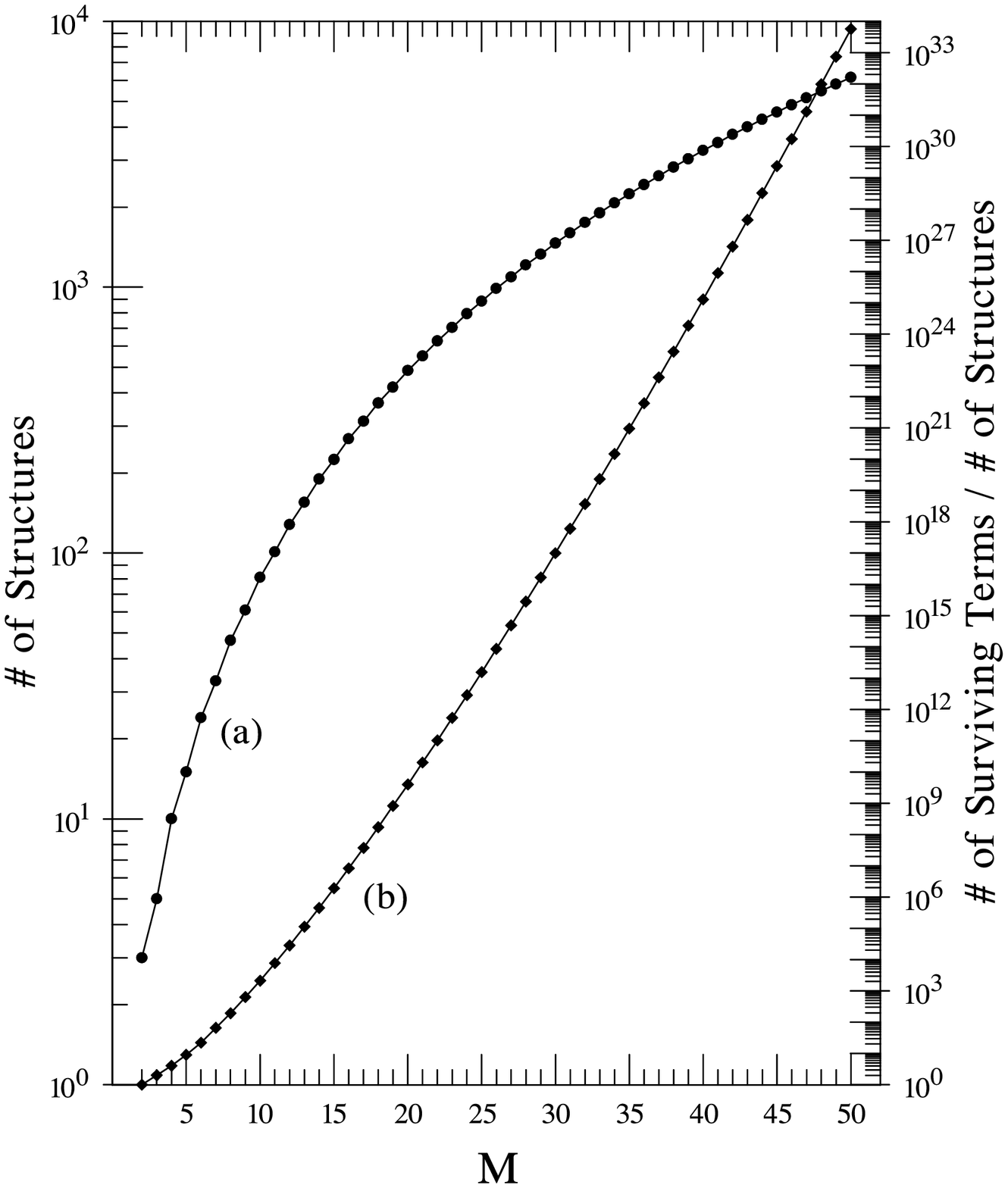}

\rm {\bf Figure 3} (a) The number of structures as function of $M$ (left
axis). (b) The average number of surviving terms per structure as function of
$M$ (right axis).
\end{center}

\newpage
\begin{center}
\includegraphics[scale=0.87,angle=0]{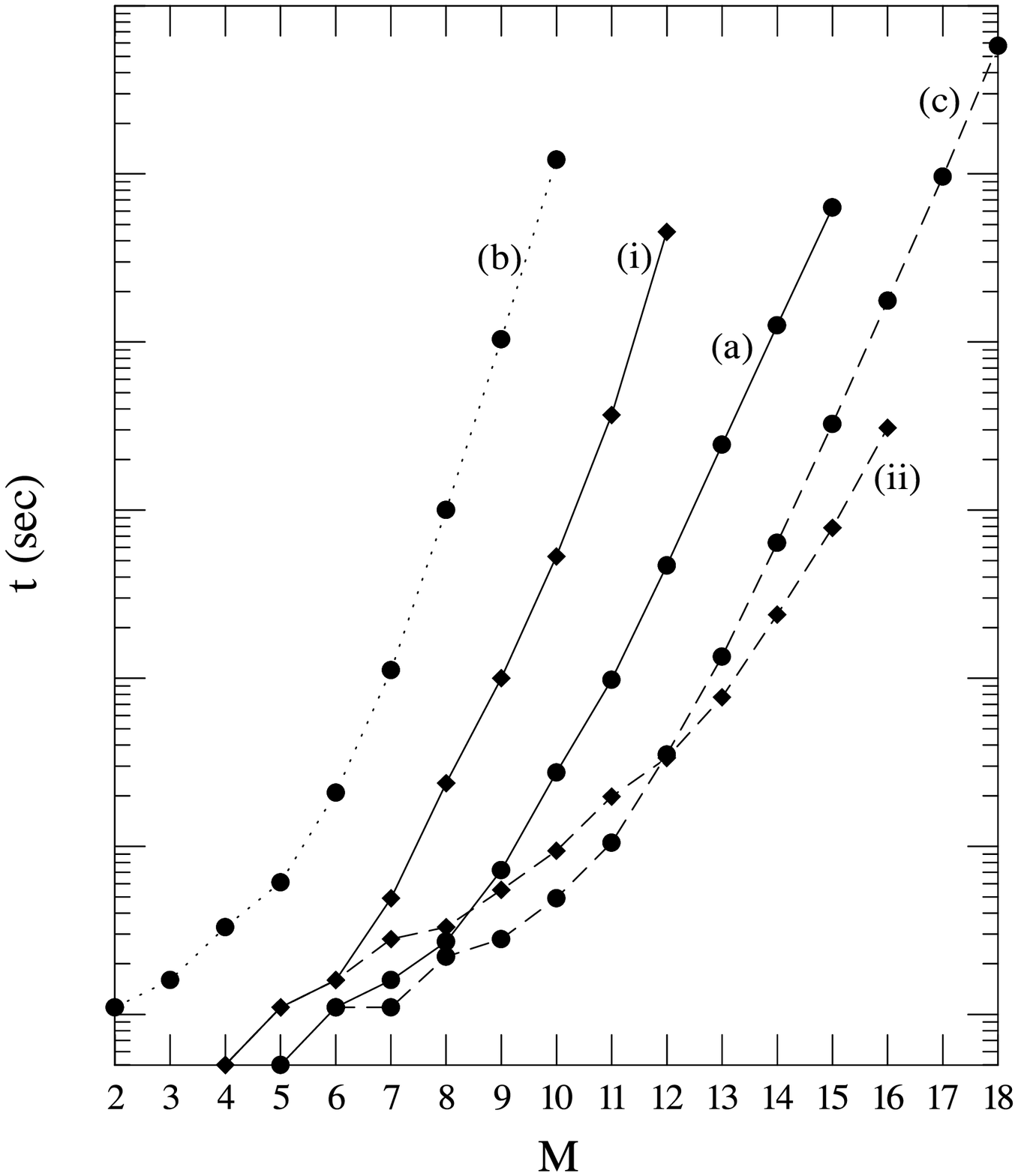}

\rm {\bf Figure 4} Curves (a)-(c) depict the consumed time for the
completion of subroutine MULTIPLYALL as function of $M$.
(a): Our new program with the output stored in files,
(b): Our old program [13] with the output stored in files,
(c): Our new program with suppressed output.
Curves (i)-(ii) depict the consumed time for the
completion of subroutine TRACE as function of $M$, for our new
program.
(i): Full result, (ii): Result with only the first term of each structure.
(Time in all cases in measured using as processor INTEL CELERON at 1.8 GHz).
\end{center}

\newpage
\begin{center}
\includegraphics[scale=0.93,angle=0]{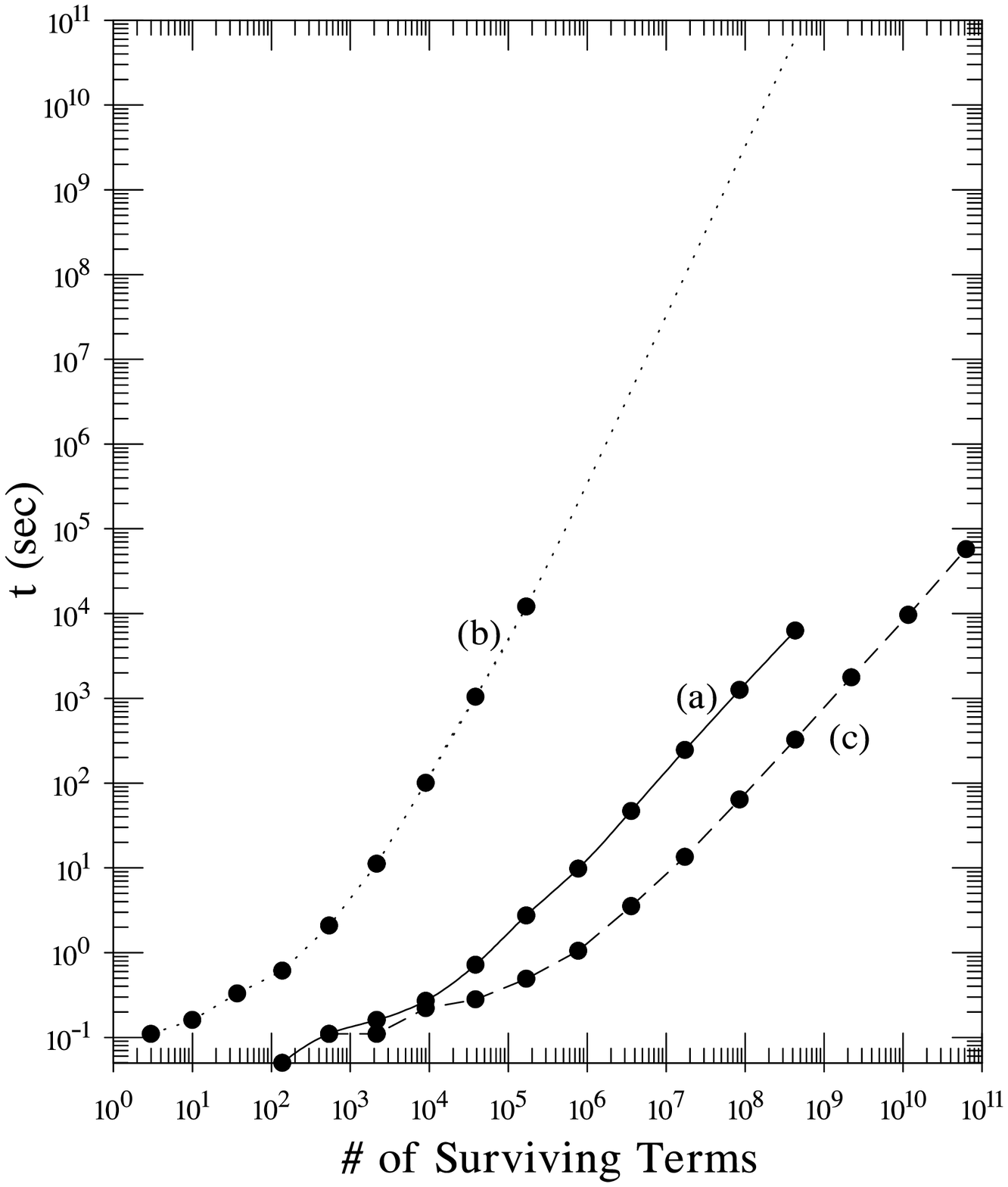}

\rm {\bf Figure 5} The consumed time in subroutine MULTIPLYALL
((a)-(c)) and TRACE ((i)-(ii)), as in Fig.~4, as function of the
surviving terms in the final result. Curve (a) has been fitted with by
a linear function and (b) has been extrapolated up to the surviving terms
for $M=15$ by a quadratic function.
\end{center}

\end{document}